\documentclass[12pt]{article}
\usepackage[reqno]{amsmath}
\usepackage{epsfig}
\usepackage{array}
\usepackage{float}

\usepackage{a4}
\usepackage{epsfig}
\usepackage{a4wide}
\usepackage{wasysym}

\def\Jo#1#2#3#4{{\it #1} {\bf #2}, #3 (#4)}

\def\NPB{{Nucl. Phys.} {\bf B}}

\def\PLB{{Phys. Lett.}  {\bf B}}
\def\PRL{Phys. Rev. Lett.}
\def\PRD{{Phys. Rev.} {\bf D}}

\def\ZPC{{Z. Phys.} {\bf C}}

\def\JHEP{JHEP}

\def\ra{\rightarrow}
\def\be{\begin{equation}}
\def\ee{\end{equation}}
\def\gs{\mathrel{
   \rlap{\raise 0.511ex \hbox{$>$}}{\lower 0.511ex \hbox{$\sim$}}}}
\def\ls{\mathrel{
   \rlap{\raise 0.511ex \hbox{$<$}}{\lower 0.511ex \hbox{$\sim$}}}}

\newcommand{\ba}{\begin{array}{c}}
\newcommand{\baz}{\begin{array}{cc}}
\newcommand{\bad}{\begin{array}{ccc}}
\newcommand{\bea}{\begin{equation} \begin{array}{c}}
\newcommand{\eea}{ \end{array} \end{equation}}
\newcommand{\ea}{\end{array}}
\newcommand{\D}{\displaystyle}

\newcommand{\meff}{\mbox{$\langle m \rangle$}}

\def\ra{\rightarrow}

\def\gtap{\mathrel{
   \rlap{\raise 0.511ex \hbox{$>$}}{\lower 0.511ex \hbox{$\sim$}}}}
\def\ltap{\mathrel{
   \rlap{\raise 0.511ex \hbox{$<$}}{\lower 0.511ex \hbox{$\sim$}}}}

\newcommand{\betabeta}{\mbox{$(\beta \beta)_{0 \nu}  $}}

\begin{document}

\title{
\hfill {\small Ref. SISSA 9/2003/EP}\\
\vspace{-0.3cm}
\hfill {\small UCLA/03/TEP/04}\\ %
\vspace{-0.3cm}
\hfill {\small hep-ph/0302054} \\ \vskip 1cm
\bf
On the Connection of Leptogenesis with Low Energy $CP$ Violation and
LFV Charged Lepton Decays}
\author{S. Pascoli$^a$,
S.T.\ Petcov\footnote{Also at: Institute of Nuclear Research and
Nuclear Energy, Bulgarian Academy of Sciences,
1784 Sofia, Bulgaria}$\, \,^{b,c}$
and
W.\ Rodejohann$^b$\\ \\
$^a${\normalsize \it Department of Physics, University of California,
Los Angeles CA 90095-1547, USA}\\[0.3cm]
$^b${\normalsize \it Scuola Internazionale Superiore di Studi Avanzati,
I-34014 Trieste, Italy}\\[0.3cm]
$^c${\normalsize \it Istituto Nazionale di Fisica Nucleare,
Sezione di Trieste, I-34014 Trieste, Italy}\\}
\date{}
\maketitle
\thispagestyle{empty}
\begin{abstract}
Assuming only a hierarchical structure of
the heavy Majorana neutrino
masses and of the neutrino Dirac mass matrix
$m_D$ of the see--saw mechanism, we find that
in order to produce
the observed baryon asymmetry of the Universe via 
leptogenesis,
the scale of $m_D$ should be given by the 
up--quark masses. 
Lepton flavor violating decays $\mu \rightarrow e + \gamma$,
$\tau \rightarrow \mu + \gamma$ and 
$\tau \rightarrow e + \gamma$
are considered and a characteristic 
relation between their decay rates 
is predicted. 
The effective Majorana mass
in neutrinoless double beta decay 
depends on the $CP$ violating phase 
controlling the leptogenesis 
if one of the heavy Majorana neutrinos 
is much heavier than the other two. 
Successful leptogenesis requires 
a rather mild mass hierarchy between the latter. 
The indicated hierarchical relations 
are also compatible
with the low-energy neutrino 
mixing phenomenology.
 The scenario under study is compatible
with the low--energy neutrino 
mixing phenomenology.
The $CP$ violation effects in neutrino 
oscillations can be observable.
In general, there is no direct connection 
between the latter and the $CP$ violation
in leptogenesis.
If the $CP$ violating phases
of the see--saw model satisfy 
certain relations,
the baryon asymmetry 
of the Universe and 
the rephasing invariant
$J_{CP}$ which determines 
the magnitude of the $CP$ violation
effects in neutrino oscillations,
depend on the same $CP$ violating phase and
their signs are correlated. 

\end{abstract}

\newpage

\section{\label{sec:intro}Introduction}

Explaining both the observed 
baryon asymmetry of the Universe ($Y_B$) and 
the smallness of the neutrino masses, can be done 
successfully by combining the see--saw \cite{seesaw} and the 
leptogenesis \cite{lepto} mechanisms.
The first predicts the neutrinos to be Majorana particles
and introduces additional right--handed  
heavy Majorana neutrinos, whose 
out--of--equilibrium decay in the early Universe 
generates $Y_B$ via the second. 
The mounting evidence in 
favor of neutrino oscillations 
(see, e.g., \cite{SNO2,KamLAND,fogli,SKatm00})
gave analyzes of these two mechanisms a firmer 
phenomenological basis.   
Establishing a connection between the low 
energy (neutrino mixing) 
and high energy (leptogenesis) parameters 
has gathered much attention in recent years 
\cite{comb,ich,others,others2,nolowCP,23,231,triangular,lissac}. 
However, the number of parameters 
of the see--saw mechanism is 
significantly larger than 
the number of quantities measurable 
in the ``low energy'' neutrino experiments.
It is the neutrino Dirac mass matrix $m_D$ that 
contains most of the unknown parameters 
and on whose knowledge any statement
about a possible connection
of low and high energy phenomena relies. 
While, in general, it is impossible to
establish a direct link between the  
phenomena related to neutrino 
mixing and to leptogenesis,
most specific models usually allow 
very well for such a connection. \\ 

   In the present article we investigate
the possible link between the leptogenesis,
taking place at ``high energy'', and
the ``low energy'' phenomena associated
with the existence of neutrino mixing.
Our analysis is based on the assumption    
of a hierarchical structure of the 
heavy Majorana neutrino masses and 
of the Dirac mass matrix $m_D$, 
both of which are part of the see--saw mechanism. 
As specific examples 
of low energy phenomena 
related to neutrino mixing
we consider the lepton flavour violating
(LFV) charged lepton decays,
$\mu \rightarrow e + \gamma$, 
$\tau \rightarrow \mu + \gamma$ and
$\tau \rightarrow e + \gamma$,
the effective Majorana mass 
in neutrinoless double beta (\betabeta--) decay, 
$|\meff|$, and the $CP$ violating asymmetry in 
neutrino oscillations.
We analyze, in particular, the
effects of the high energy 
$CP$ violating phases, which 
control the generation of 
the baryon asymmetry in the 
leptogenesis scenario,
in $|\meff|$ and in 
the leptonic $CP$ violation 
rephasing invariant 
$J_{CP}$ \cite{PKSP88}
which determines the magnitude of the
$CP$ violation in neutrino oscillations 
\footnote{For similar earlier attempts (not taking 
into account the questions of 
$CP$ violation and LFV decays),
see, e.g., \cite{first,berger}.}.\\

  In the next Section we briefly review 
the formalisms associated with the see--saw
mechanism and with the leptogenesis scenario, 
and that of LFV charged lepton decays. 
Different parameterizations of $m_D$ 
used in the literature and 
the corresponding connection they imply 
between the high and low energy physical parameters are 
discussed in Section \ref{sec:how}. 
In Section \ref{sec:biun}
we formulate the assumptions
of a hierarchical structure of the 
heavy Majorana neutrino masses and 
of the Dirac mass matrix $m_D$
on which our investigation is based, 
and analyze the leptogenesis in the
so--called ``bi--unitary'' parametrization 
of $m_D$. Predictions 
for the branching rations of the 
LFV charged lepton decays are also given. 
In Section \ref{sec:connection?}   
the effects of the high energy 
leptogenesis $CP$ violating phases 
on $|\meff|$ and on 
the leptonic $CP$ violation 
rephasing invariant $J_{CP}$. 
Finally, we conclude 
in Section \ref{sec:concl}.

\section{\label{sec:frame}Framework}

The neutrino oscillation data 
can consistently be described within a
3--neutrino mixing scheme with massive
Majorana neutrinos, in which the 
light neutrino Majorana mass matrix 
$m_\nu$ is given by: 
\begin{equation} \label{eq:mnu}
m_\nu = U_{\rm PMNS} \, \, m_\nu^{\rm diag} \, \, U_{\rm PMNS}^T~.
\end{equation}
%
Here $m_\nu^{\rm diag}$ is a diagonal 
matrix containing the masses $m_{1,2,3}$
of the three massive Majorana neutrinos and 
$U_{\rm PMNS}$ is the unitary 
Pontecorvo--Maki--Nagakawa--Sakata 
\cite{PMNS} lepton 
mixing matrix, which can be parametrized as 
\begin{equation}
\begin{array}{c}
\label{eq:Upara}
U_{\rm PMNS} = V  \; 
{\rm diag}(1, e^{i \alpha}, e^{i (\beta + \delta)}) \\[0.3cm]
= \left( \bad 
c_1 c_3 & s_1 c_3 & s_3 e^{-i \delta} \\[0.2cm] 
-s_1 c_2 - c_1 s_2 s_3 e^{i \delta} 
& c_1 c_2 - s_1 s_2 s_3 e^{i \delta} 
& s_2 c_3 \\[0.2cm] 
s_1 s_2 - c_1 c_2 s_3 e^{i \delta} &
- c_1 s_2 - s_1 c_2 s_3 e^{i \delta} 
& c_2 c_3\\ 
               \ea   \right) 
 {\rm diag}(1, e^{i \alpha}, e^{i (\beta + \delta)}) \, , 
\end{array}
\end{equation}
%
where $\delta$ is a Dirac $CP$ violating phase, 
$\alpha$ and $\beta$ are Majorana 
$CP$ violating phases 
\cite{BHP80,Doi81},
$c_i = \cos\theta_i$ and $s_i = \sin\theta_i$. 
The angles $\theta_1$ and $\theta_2$ control the oscillations 
of solar and atmospheric neutrinos, respectively. The angle 
$\theta_3$ is limited by the CHOOZ and Palo Verde 
reactor $\bar{\nu}_e$ experiments:
one has $\sin^2 \theta_{3} < 0.05$ \cite{CHOOZ,fogli}. 
The Dirac phase $\delta$ can be measured, in principle,
in long base--line neutrino oscillation experiments 
(see, e.g., \cite{AMMS,CPosc}). The flavour neutrino oscillations 
are insensitive to the Majorana phases 
$\alpha$ and $\beta$ \cite{BHP80,Lang87}. 
Information about these phases can be obtained
by studying processes in 
which the total lepton charge $L$
is not conserved and changes by two units 
\cite{BGKP96,BPP1,PPR1,ichJPG,kayser}: \betabeta--decay, 
$K^{+} \rightarrow \pi^{-} + \mu^{+} + \mu^{+}$, etc.\\ 

   Introducing a Dirac neutrino mass 
term and a Majorana mass
term for the right--handed   
neutrinos via the Lagrangian 
\begin{equation} \label{eq:L}
 -{\cal{L}} = \overline{\nu_{Li}} \, (m_D)_{ij} \, N_{Rj} \, + 
\frac{1}{2} \, \overline{(N_{Ri})^c} \, (M_R)_{ij} \, N_{Rj} ~, 
\end{equation}
%
leads for sufficiently large $M_R$ 
to the well know see--saw \cite{seesaw} formula 
\begin{equation} \label{eq:seesaw}
m_\nu \simeq - m_D \, M_R^{-1} \, m_D^T~, 
\end{equation}
%
where terms of order  ${\cal O}(M_R^{-2})$ 
are neglected.
In order to explain the smallness 
of neutrino masses, the see--saw 
mechanism requires the existence of 
heavy right--handed Majorana 
neutrinos. The latter 
can create a lepton asymmetry 
via their $CP$ violating
out--of--equilibrium decays induced by the see--saw related 
Yukawa couplings. 
At later epoch the lepton asymmetry
is converted into the 
baryon asymmetry of the Universe 
through sphaleron mediated
processes \cite{lepto}. 
Thus, leptogenesis is naturally 
incorporated in the see--saw model,
which makes the model particularly 
attractive.

\subsection{\label{sec:leptogenesis}Leptogenesis}
Leptogenesis fulfills all of Sakharov's 
three conditions \cite{sakharov} 
for generation of a non--vanishing 
baryon asymmetry $Y_B$. The 
requisite $CP$ violating asymmetry 
is caused by the interference of 
the tree level contribution and 
the one--loop corrections in the decay rate
of the lightest of the three heavy 
Majorana neutrinos, 
$N_1 \ra \Phi^- \, \ell^+$ and $N_1 \ra \Phi^+ \, \ell^-$:
\begin{equation} \label{eq:eps}
\ba 
\varepsilon_1 = \frac{\D \Gamma (N_1 \ra \Phi^- \, \ell^+) - 
\Gamma (N_1 \ra \Phi^+ \, \ell^-)}{\D \Gamma (N_1 \ra \Phi^- \, \ell^+) +  
\Gamma (N_1 \ra \Phi^+ \, \ell^-)} \\[0.4cm]
\simeq  \D \frac{\D 1}{\D 8 \, \pi \, v^2} \frac{\D 1}{(m_D^\dagger m_D)_{11}} 
\sum_{j=2,3} {\rm Im} (m_D^\dagger m_D)^2_{1j} \, \left( f(M_j^2/M_1^2) + 
g(M_j^2/M_1^2) \right)~.
\ea 
\end{equation}
%
Here $v \simeq 174$ GeV is the electroweak symmetry breaking 
scale. The function $f$ stems from vertex 
\cite{lepto,vertex} and $g$ from self--energy \cite{self} contributions: 
\begin{equation} \label{eq:fapprox}
\baz 
f(x) = \sqrt{x} \left(1 - 
(1 + x) \, \ln \left(\frac{\D 1 + x}{\D x}\right) \right) \; , \, 
& \, g(x) = \frac{\D \sqrt{x}}{\D 1 - x}
\ea \, . 
\end{equation}
%
For $x \gg 1$, i.e., for hierarchical 
heavy Majorana neutrinos, 
one has $f(x) + g(x) \simeq -\frac{3}{2\sqrt{x}}$. 
The baryon asymmetry is obtained via 
\begin{equation} \label{eq:YB}
Y_B = a \, \frac{\kappa}{g^\ast} \, \varepsilon_1~,
\end{equation}
%
where $a \simeq -1/2$ is the fraction of 
the lepton asymmetry converted into 
a baryon asymmetry \cite{conv}, 
$g^\ast \simeq 100$ is the number of massless 
degrees of freedom at the time of the decay, 
and $\kappa$ is a dilution factor that is 
obtained by solving the Boltzmann equations. 
Typically, one gets $Y_B \sim 10^{-10}$ when 
$\varepsilon_1 \sim (10^{-6} - 10^{-7})$ and 
$\kappa \sim (10^{-3} - 10^{-2})$. 
We note that this estimate of $Y_B$ 
is valid in the supersymmetric (SUSY) 
theories as well
since in the latter  
both $g^\ast$ and $\varepsilon_1$ 
are larger approximately 
by the same factor of 2. 

\subsection{\label{sec:muegsum}LFV Charged Lepton Decays}
In extensions of the Standard Theory
including massive neutrinos, 
lepton flavour non--conserving (LFV) processes 
such as $\mu\ra e + \gamma$, $\mu \ra 3 e$, 
$\tau\ra \mu+\gamma$, $\tau\ra e+\gamma$, etc.
are predicted to take place (see, e.g., \cite{BiPet87}).
However, in the non--SUSY versions 
of the see--saw model,
the decay rates and cross sections 
of the LFV processes are strongly suppressed and 
are practically unobservable~\cite{SP76,lfv}. 
The most stringent experimental limit
on the $\mu\ra e + \gamma$ 
decay branching ratio reads~\cite{mega}
\begin{equation}
BR(\mu\ra e+\gamma) < 1.2\times 10^{-11}~.
\end{equation}
%
There are prospects 
to improve this limit 
by $\sim 3$ orders of magnitude in the future (see, e.g., \cite{psi}). 
The existing bounds on the LFV decays of the $\tau-$lepton 
are considerably less stringent: one has, e.g., 
$BR(\tau \ra \mu +\gamma) < 6.0\times 10^{-7}$ \cite{taulfv}. 
There are possibilities
to reach a sensitivity to values of
$BR(\tau \ra \mu +\gamma) \sim (10^{-8} - 10^{-9})$
at B--factory and LHC experiments \cite{Ohshima}.\\ 
   
In SUSY theories 
incorporating the see--saw mechanism,
new sources of lepton flavour violation 
arise. In SUSY GUT theories these are 
typically related to
the soft SUSY--breaking terms of the Lagrangian. 
Under the assumption of flavour universality 
of the SUSY breaking sector (scalar masses 
and trilinear couplings) at the GUT scale $M_X$, 
new lepton flavour 
non--conserving couplings are induced 
at low energy by the matrix of neutrino Yukawa
couplings $m_D/v$ through
renormalization group running
of the scalar lepton masses.
The latter generates 
non--zero flavour non--diagonal entries in
the slepton mass matrices and in 
the trilinear scalar couplings.
These were shown to be 
proportional to $m_D m_D^\dagger$~\cite{borz}.
More specifically, using 
the leading--log approximation,
one can estimate ~\cite{borz,iba,JohnE} 
the branching ratio of the 
charged lepton decays 
$\ell_i \rightarrow \ell_j + \gamma$,
$\ell_i(\ell_j) = \tau, \mu, e$ for $i(j) = 3,2,1$,
$i > j$,
\begin{equation} \label{eq:muegBR}
BR(\ell_i \rightarrow \ell_j + \gamma) \simeq 
\alpha^3 \, \left( \frac{\D (3 + a_0) \, m_0^2}
{\D 8 \, \pi^2 \, m_S^4 \, G_F \, v^2} \right)^2  
\left|(m_D L m_D^\dagger)_{ij}\right|^2 \tan^2 \beta~,
\end{equation}
%
where $m_S$ denotes a slepton mass, $m_0$ is the universal 
mass scale of the sparticles and $a_0$ is a trilinear 
coupling (all at $M_X$). The diagonal matrix $L$ reads: 
$L = {\rm diag}(L_1,L_2,L_3)$, where $L_i = \log M_X/M_{i}$.    
To get a feeling for the numbers involved, for $a_0 = {\cal O}(1)$,  
$m_S \sim m_0 \sim 10^2$ GeV, $M_X \sim 10^{16}$ GeV and 
$M_{i} \sim 10^{10}$ GeV, one finds 
\begin{equation}
BR(\ell_i \rightarrow \ell_j + \gamma) \sim 10^{-15} \, \tan^2 \beta \,   
\left|\frac{(m_D m_D^\dagger)_{ij}}{\rm GeV^2} \right|^2~. 
\end{equation}
This is within the planned sensitivity of the 
next generation experiments. The predicted values 
of  $BR(\ell_i \rightarrow \ell_j + \gamma)$ are strongly 
dependent on the SUSY parameters, a topic beyond the 
scope of the present article.\\

Since the rates of LFV charged lepton decays 
depend on $m_D m_D^\dagger$, 
the decay asymmetry $\varepsilon_1$ 
depends on $m_D^\dagger m_D$ and 
leptonic $CP$ violating effects originate from 
$m_\nu \sim m_D m_D^T$, one might in 
a given model 
expect some interplay between these three phenomena. We shall 
investigate this in Sections \ref{sec:biun} and 
\ref{sec:connection?}.

\section{\label{sec:how}How it Works}
\subsection{\label{sec:para}Parameterizations}

At low energy and 
in the case of $N$ flavour neutrinos, 
we can measure, in principle, 
the $N \times N$ light neutrino  
mass matrix $m_\nu$. It contains 
$N$ mass eigenstates, $\frac{1}{2} N (N - 1)$
mixing angles and 
$\frac{1}{2} N (N - 1)$ phases, 
thus altogether $N^2$ measurable quantities. 
Of these $\frac{1}{2} N (N + 1)$ are real  
and $\frac{1}{2} N (N - 1)$ are  $CP$ violating phases. 
One can easily count the number 
of {\it all} see--saw parameters in 
the weak basis, in which both, $M_R$ and 
the charged lepton mass matrix are real and diagonal. In this basis 
the Dirac mass matrix $m_D$ contains 
all information about $CP$ violation. 
We shall 
discuss now three possible 
parameterizations of $m_D$.\\ 

\noindent {\bf 1.\ Bi--unitary parametrization}\\

With $N$ light and $N'$ heavy neutrinos,
denoting this particle content as 
the ``$N \times N'$ see--saw model'', 
one can write the complex $N \times N'$ Dirac mass matrix as 
\begin{equation} \label{eq:mdULUR}
m_D = U_L^\dagger \, m_D^{\rm diag} \, U_R~,  
\end{equation}
%
where $U_L$ ($U_R$) is a 
unitary $N \times N$ ($N' \times N'$) matrix and  
$m_D^{\rm diag}$ is a real matrix with non--zero elements only 
at its $ii$ entries. Thus, $m_D^{\rm diag}$ 
contains $\min(N,N')$ real parameters we shall denote 
by $m_{Di}$.  
Any unitary $N \times N$ matrix can be written as 
\begin{equation}
U = e^{i \Phi} \, P \, 
\tilde{U} \, Q~, 
\end{equation}
%
where $P \equiv {\rm diag} (1,e^{i \phi_1},e^{i \phi_2},\ldots)$
and $Q \equiv {\rm diag} (1,e^{i \alpha},e^{i \beta},\ldots) $ 
are diagonal phase matrices with 
$(N - 1)$ phases, and $\tilde{U}$ is a unitary 
``CKM--like'' matrix containing 
$\frac{1}{2} (N - 2) (N - 1)$ phases 
and $\frac{1}{2} N (N - 1)$ angles. 
In total $U$ contains 
$\frac{1}{2} N (N + 1)$ phases. 
Parameterizing $U_L^\dagger$ and $U_R$ in the same way, 
we get for the Dirac mass matrix 
\begin{equation} \label{eq:mdULUR2}
m_D = e^{i(\Phi_R + \Phi_L)} \, P_L\,
 \tilde{U}_L \, Q_L\,  
m_D^{\rm diag} \, P_R \, \tilde{U}_R \, Q_R~,  
\end{equation}
%
where the index $L(R)$ indicates that the 
angles and phases in 
the respective matrices
also carry this index. The common phase 
($\Phi_R + \Phi_L $) 
and the $(N - 1)$ phases in $P_L$ can be absorbed by 
a redefinition of the charged lepton fields. 
>From the original $(N + N' - 2)$ phases in 
the matrices $Q_L$ and $P_R$, only 
$(N + N' - 2) - (\max(N,N') - 1)$  
appear in the product $Q_L \, m_D^{\rm diag} \, P_R$. 
Thus, we can write 
\begin{equation}
m_D = \tilde{U}_L  \, 
W \, m_D^{\rm diag} \, \tilde{U}_R \, Q_R~,
\end{equation} 
%
where $W$ is a diagonal matrix containing 
$N' - 1$ ($N - 1$) phases if $N > N'~(N \leq N')$.  
The number of physical $CP$ violating phases is therefore
\begin{equation}
\begin{array}{c}  
\frac{1}{2} N (N + 1) + \frac{1}{2} N' (N' + 1) - 2 - (N - 1) - 
(\max(N,N') - 1)\\[0.3cm]
 = \frac{1}{2} N (N - 1) +  \frac{1}{2} N' (N' + 1) - 
\max(N,N')~\\[0.3cm] 
= \frac{1}{2} N (N - 3) +  \frac{1}{2} N' (N' - 3) 
+ N' + \min(N,N')  
\end{array}~.
\end{equation}
%
Adding the number of the real parameters $m_{Di}$, $\min(N,N')$, 
of the masses $M_{i}$, $N'$, and 
of the angles in $\tilde{U}_L$  
$(\tilde{U}_R)$, $\frac{1}{2} N (N - 1)$
$(\frac{1}{2} N' (N' - 1))$, we 
obtain the total number of 
parameters in the $N \times N'$ see--saw model. 
The result is $N (N - 1) + N' (N' + 1) - \max(N,N') 
+ \min(N,N')$. For $N = N'$ this reduces to $2 N^2$ parameters, 
$N (N - 1)$ phases and $N (N + 1)$ real one. 
In general, the difference between the number of real parameters 
and phases is $N + N'$.\\

  Comparing the number of low and 
high energy parameters, we see that 
integrating out the heavy Majorana neutrinos 
leave us short of a total of 
$N' (N' + 1) - N - \max(N,N') + \min(N,N')$ parameters, 
of which 
$\frac{1}{2} N' (N' + 1) - N + \min(N,N')$ are real and 
$\frac{1}{2} N' (N' + 1) - \max(N,N')$ are phases. For $N = N'$ 
half of the parameters ``get lost'', $\frac{1}{2} N (N - 1)$ phases 
and $\frac{1}{2} N (N - 1)$ real ones. 
  The counting of the independent 
physical parameters in $m_D$
through the independent parameters
in $U_{L,R}$ made
above, is valid for $|N - N'| = 0,1$. For 
arbitrary $|N - N'| > 1$, the total number
of independent physical parameters in
$m_D$ is obviously $N (2 N' - 1)$, 
of which $NN'$ are moduli and 
$N(N' - 1)$ are phases.
Their number is smaller than the number 
our counting through $U_{L,R}$ gives
because when $|N - N'| > 1$ there are more
unphysical parameters in $U_{L,R}$.\\

    The usual $3 \times 3$ see--saw model 
has therefore 18 parameters, which are composed of  
12 real parameters and 6 phases. 
Integrating out the heavy Majorana neutrinos 
leaves us with the observable 
mass matrix $m_\nu$, which 
contains 9 observable parameters ---  
6 real and 3 phases. Hence, half of 
the parameters of the model get ``lost''  
at low energy. However, 
in approaches to reconstruct the 
high energy physics from low energy data, 
one usually 
assumes the existence 
of relations between mass matrices 
(e.g.,\ $m_D=m_{\rm up}$), 
which reduces the number of unknown parameters. 
If in addition specific textures 
in the matrices are implemented, 
the situation improves further.\\ 

   It is worth noting that ``reduced'' 
models allow for a simpler connection between 
the high and low energy physics phenomena.
Consider the minimal $3\times2$ 
see--saw model \cite{23,231}, which 
contains only 2 heavy Majorana neutrinos. 
This model has altogether 11 parameters, 
8 real and 3 $CP$ violating 
phases. Comparing with the 9 observable 
parameters in $m_\nu$, we see that only 2 real
parameters are ``missing'' at low energy. 
Interestingly, this makes the model 
superior to the even more reduced 
``too minimal'' $2\times2$ see--saw model with 
2 heavy and 2 light Majorana neutrinos 
\footnote{This model can be motivated by a 
vanishing mixing angle $\theta_3$. The $\nu_e$  
oscillates then into the flavor 
state $(\nu_\mu + \nu_\tau)/\sqrt{2}$
with an amplitude $\sin^2 2 \theta_\odot$.}. 
The latter has 8 parameters, 
two of which are phases, and this has to be 
compared with the 4 parameters (including 1 phase)
a $2\times2$ light Majorana neutrino mass
matrix effectively contains.
 
  We shall use in the analysis that follows 
the bi--unitary parametrization. 
Before proceeding further, however, 
we will describe briefly 
two alternative parameterizations.\\

\noindent {\bf 2.\ Triangular parametrization}\\
This parametrization relies on 
the property that any complex matrix 
can be written as a product of a unitary and a lower triangular
matrix \cite{triangular}: 
\begin{equation} \label{eq:mdtri}
m_D = v \, U \, Y~. 
\end{equation}
%
The triangular matrix $Y$ has zeros 
as entries above the diagonal axis 
and contains  3 phases in the 
three off--diagonal entries. The 
unitary matrix $U$ can be parametrized 
in analogy to the PMNS 
matrix, i.e., in the form of a CKM--like 
matrix with one phase times a 
diagonal matrix containing the other 
two phases. 
An analysis of leptogenesis using this 
parametrization was performed in Ref.\ \cite{lissac}.\\

\noindent {\bf 3.\ Orthogonal parametrization}\\
The following parametrization shows 
clearly that without any assumptions 
there is no connection between the low 
and high energy parameters governing 
respectively neutrino mixing and leptogenesis. One can write
the Dirac matrix as \cite{iba}  
\begin{equation} \label{eq:mdO}
m_D =  i \, U_{\rm PMNS} \, \sqrt{m_\nu^{\rm diag}} \, R \, \sqrt{M_R}~,
\end{equation}
%
where $R$ is a complex orthogonal matrix. 
It contains 3 real parameters 
and 3 phases, which, together with the 3 $M_{ i}$ and the 9 
parameters from $m_\nu$, sums up again to a total of 18 parameters. 
In Eq.\ (\ref{eq:mdO}), $m_D$ seems to contain all 18 parameters.
However, the product
$\sqrt{m_\nu^{\rm diag}} \, R \, \sqrt{M_R}$ includes only
9 independent parameters.


\subsection{\label{sec:lepto}Leptogenesis, $m_\nu$ and 
LFV Charged Lepton Decays}


The decay asymmetry $\varepsilon_1$ 
depends on the three mass eigenvalues 
$M_{ i}$ and on the hermitian matrix 
$m_D^\dagger m_D$. Due to its 
hermiticity, the latter will 
have a reduced number of parameters
with respect to $m_D$:
for every row (or column) we get rid of one 
complex number, or equivalently of one 
real parameter and one phase.  
Indeed, 
\begin{equation} \label{eq:mddmd}
m_D^\dagger m_D = 
\left\{ 
\baz 
U_R^\dagger \, (m_D^{\rm diag})^2 \, U_R~, & \mbox{bi--unitary;} \\[0.3cm]
v^2 \, Y^\dagger \, Y~,              & \rm triangular; \\[0.3cm]
 \sqrt{M_R} \, R^\dagger \, m_\nu^{\rm diag} \, R \, \sqrt{M_R}~, 
& \rm orthogonal, \\
\ea 
\right. 
\end{equation}
%
depends on $15 - 6 = 9$ parameters, which 
for the bi--unitary parametrization
are the 3 angles and 3 phases in 
$U_R$ and the 3 mass eigenvalues $m_{Di}$;
for the triangular parametrization,
these are the 6 real entries and 3 phases in $Y$.
In the orthogonal parametrization 
we have the 9 parameters from the 
combination $\sqrt{m_\nu^{\rm diag}} \, R \, \sqrt{M_R}$. 
Of course, there are three leptogenesis phases in each case.\\

\begin{figure}
\begin{center}
\epsfig{file=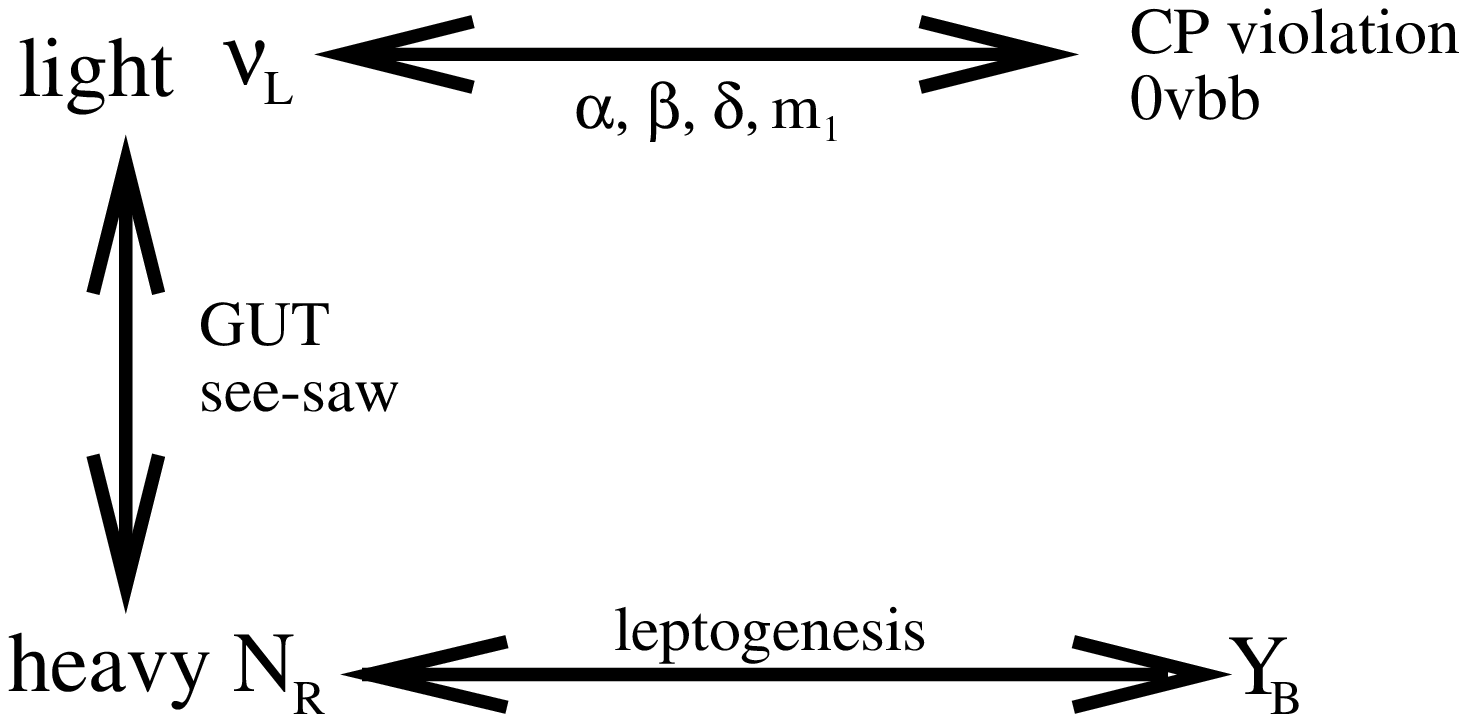,width=15cm,height=8cm}
\caption{\label{fig:principle}Connection between low energy lepton number and 
$CP$ violation with the baryon asymmetry $Y_B$ via the leptogenesis mechanism. 
Without the left vertical arrow there is none.}
\end{center}
\end{figure}

    The form of $m_D^\dagger m_D$ in 
the orthogonal parametrization 
shows that the PMNS matrix $U_{\rm PMNS}$ does not 
enter into the expression for 
$\varepsilon_1$ and that there 
is, in general, no connection 
between the low energy 
observables in $U_{\rm PMNS}$ and leptogenesis. 
In particular, there is no connection 
between the low and high energy $CP$ violation. 
Figure \ref{fig:principle} shows a 
sketch of the situation: in order 
to connect low energy lepton charge 
non--conservation and $CP$  
violation effects (making 
an assumption
about the currently unknown 
values of the phases and the lowest 
neutrino mass eigenstate) 
with the baryon asymmetry
of the Universe $Y_B$, 
there must exist 
a connection between
the light left--handed neutrinos $\nu_L$ 
and the heavy right--handed Majorana 
neutrinos $N_R$.  
This typically involves some kind of see--saw 
mechanism within the framework
of a GUT theory. The $N_R$ produce $Y_B$ 
via the leptogenesis mechanism. Without 
the crucial GUT or see--saw link 
there is no connection.\\

   Inserting the form of $m_D$ from 
Eqs.\ (\ref{eq:mdULUR}) and (\ref{eq:mdtri}) in 
Eq.\ (\ref{eq:seesaw}) shows 
\footnote{Note that inserting Eq.\ (\ref{eq:mdO}) in (\ref{eq:seesaw}) 
yields an identity.} 
that all six phases in $m_D$  
contribute to the phases in $U_{\rm PMNS}$:
\begin{equation} \label{eq:mnupara}
m_\nu = 
\left\{ 
\baz 
- U_L^\dagger \, m_D^{\rm diag} \, U_R \, M_R^{-1} \,  U_R^T \, 
m_D^{\rm diag} \, U_L^\ast~,
& \mbox{bi--unitary;} \\[0.3cm]
-v^2 \, U \, Y \, M_R^{-1} \, Y^T \, U^T~, & \rm triangular. \\
\ea 
\right. 
\end{equation}
%
One may expect that a 
connection between the high and low 
energy $CP$ violation would exist if 
three of the phases in $m_D$ are negligible, 
so that the numbers of the $CP$ violating 
phases at high and low energy 
coincide. In order to have successful 
leptogenesis, the negligible phases 
should not be those appearing 
in $Y$ or $U_R$. Thus, one can assume that, 
e.g.,  $U$ in Eq.\ (\ref{eq:mdtri})  
is real, as has been done in \cite{lissac}. 
We shall pursue a different approach and
consider the bi--unitary parametrization 
keeping all the phases.\\

  In supersymmetric versions of the see--saw mechanism, 
the rates of LFV charged lepton decays, such as $\mu \ra e + \gamma$, 
and the $T$ violating asymmetries in, e.g.,  $\mu \ra 3e$ decay, 
depend approximately
\footnote{The actual dependence is a 
bit more evolved, see Section \ref{sec:muegsum}.}      
on \cite{borz,iba,JohnE} 
\begin{equation} \label{eq:mdmdd}
m_D m_D^\dagger = 
\left\{ 
\baz 
U_L^\dagger \, (m_D^{\rm diag})^2 \, U_L~, & \mbox{bi--unitary;} \\[0.3cm]
v^2 \, U \, Y \, Y^\dagger \, U^\dagger~,  & \rm triangular; \\[0.3cm]
 U_{\rm PMNS} \, 
\sqrt{m_\nu^{\rm diag}} \, R \, M_R \, R^\dagger \, \sqrt{m_\nu^{\rm diag}} 
\, U_{\rm PMNS}^\dagger~, 
& \rm orthogonal. \\
\ea 
\right. 
\end{equation}
%
In principle, taking also a possible data on 
the electric dipole moments of
charged leptons into account,
there are enough observables for a full 
reconstruction of the see--saw model. The bi--unitary 
parametrization seems to be especially 
convenient for this purpose. 


\section{\label{sec:biun} The bi--Unitary Parametrization for $m_D$}


As we saw, 
the $3 \times 3$ see--saw model 
has 6 phases as possible 
sources of leptonic $CP$ violation.
The 3 low energy 
$CP$ violating phases depend, in general, on all 6 see--saw 
phases. The unitary matrices 
$U_L$ and $U_R$ have a simple 
interpretation: 
$U_L$ diagonalizes $m_D m_D^\dagger$, 
thus being responsible for 
lepton flavor violation, and 
$U_R$ diagonalizes $m_D^\dagger m_D$, therefore 
being responsible for the 
total lepton number 
non--conservation. 
Since the latter is 
a necessary ingredient of leptogenesis and leads also to 
\betabeta--decay, one might assume naively 
that there will be a certain connection 
between the \betabeta--decay rate and $Y_B$. 
We shall see that within certain plausible 
assumptions this is indeed the case.\\ 

   The first step is to determine the form of 
$U_L$ and $U_R$. Since in any GUT theory  
$m_D$ is related to 
the known charged fermion masses, we can assume a hierarchical 
structure of $m_D$, i.e., $m_{D3} \gg m_{D2} \gg m_{D1}$, such that 
\begin{equation} \label{eq:mdass}
m_D \simeq {\rm diag}(m_{D1} , m_{D2}, m_{D3}) + 
{\cal{O}}({\rm few} \, \%)~,
\end{equation}
%
where the second term indicates that there are 
corrections not exceeding 10 $\%$ on both the 
diagonal and off--diagonal entries of $m_D$. 
It is helpful to parametrize $U_L^\dagger$ 
and $U_R$ in analogy to $U_{\rm PMNS}$ 
in Eq.\ (\ref{eq:Upara}), with the angles 
$\theta_i$ replaced by 
$\theta_{L(R)i}$. 
In what regards the $CP$ violating phases, we shall denote the phases in 
 $\tilde{U}_R$ and $\tilde{U}_L^\dagger$ with $\delta_R$
and $\delta_L$ respectively, the two ``Majorana phases''  
in $Q_R$ by $\alpha_R$, $\beta_R + \delta_R$ and 
the two phases in $W$ by $\alpha_W$ and $\beta_W+ \delta_L$. 
In order to obtain a hierarchical $m_D$ 
we shall assume that 
$s_{L(R)1} \sim 10^{-1} > s_{L(R)2} \sim 10^{-2} >  
s_{L(R)3}$ with $s_{L(R)3} \ls 10^{-3}$, as well as that 
$m_{D3} \gg m_{D2} \gg m_{D1}$. 
These assumptions are 
inspired by observed mixing in the quark 
sector and by the known hierarchies
between the masses of the up--type
quarks, between the masses 
of the down--type quarks and between
the charged lepton masses. 
In the numerical estimates we give in what follows
we will always use the values 
$m_{D1} \sim 100$ MeV,  $m_{D2} \sim 1$ GeV and  $m_{D3}\sim 100$ GeV,
which to a certain degree are suggested by the up--quark mass values.\\ 
 
  Once $U_L^\dagger$ ($U_R$) is chosen to contain 
hierarchical mixing angles, the requirement that $m_D$ takes the 
form (\ref{eq:mdass}) implies hierarchical mixing angles also for 
$U_R$ ($U_L^\dagger$).  
The structure of the Dirac matrix is 
then found to be  
\begin{equation} \label{eq:mdapp}
\begin{small}
m_D \simeq 
\left( 
\bad
m_{D1} - \tilde{m}_{D2} \, s_{1L} \, s_{1R} & 
e^{i \alpha_R} \, \tilde{m}_{D2} \,  s_{1L} & 
e^{i (\beta_R + \delta_R - \delta_L)} \, 
\tilde{m}_{D3} \, s_{3L}   \\[0.3cm]
- \tilde{m}_{D2} \, s_{1R} &
e^{i \alpha_R} \, \tilde{m}_{D2} & 
e^{i (\beta_R + \delta_R)} \, \tilde{m}_{D3} \, s_{2L} \\[0.3cm]
\tilde{m}_{D3} \left( s_{1R} \, s_{2R} - e^{i \delta_R} \, s_{3R} \right)  
& - e^{i \alpha_R } \, \tilde{m}_{D3} \, s_{2R} 
&  e^{i (\beta_R + \delta_R)} \, \tilde{m}_{D3} 
\ea
\right)~,
\end{small}
\end{equation}
%
where 
$\tilde{m}_{D2} \equiv m_{D2} \, e^{i \alpha_W}$,
$\tilde{m}_{D3} \equiv m_{D3} \, e^{i (\beta_W + \delta_L) }$ and 
we have set all the $\cos \theta_{L(R)i}$ to one. 
Products of the $\sin\theta_{L(R)i}$  
which are smaller than 
$10^{-3}$ are neglected.
We have neglected also terms of order 
${\cal O}( m_{D3} \, s_{2L} \, s_{2R})$ with respect to
$m_{D2}$. 
Not surprisingly, the off--diagonal  
entries of $m_D$ are suppressed with respect 
to the diagonal terms by the small 
quantities $s_{iL(R)}$, $i = 1,2,3$. A hierarchical 
structure of $m_D$ is most naturally described using the 
bi--unitary parametrization.\\  

Rather than tuning the 
parameters involved 
in the bi--unitary parametrization
to reproduce precisely the 
neutrino mass squared 
differences and mixing angles
determined from the 
available neutrino oscillation data,  
we shall focus only on the 
dependence 
of the quantities of interest 
on the parameters in $m_D$ and $M_R$: 
we are primarily interested 
to find out under which 
circumstances a connection between 
the neutrino mixing related phenomena
and leptogenesis is possible. 
We made nevertheless  
extensive consistency checks regarding the 
form of $m_\nu$ as obtained with 
the matrix $m_D$ in Eq.\ (\ref{eq:mdapp}) and with $M_R$.
Using the form of $m_D$ given in Eq.~(\ref{eq:mdapp})
and assuming  a hierarchy for $M_i$, $M_1 \ll M_2 \ll M_3$,
from Eq.~(\ref{eq:seesaw}) one 
can obtain a theoretically predicted expression 
for the low energy neutrino mass matrix, $m_\nu^\mathrm{th}$. 
The form of $m_\nu^\mathrm{th}$ 
should be confronted with 
the neutrino mass matrix which is obtained
from Eq.~(\ref{eq:mnu}), $m_\nu^\mathrm{exp}$.
For the angles $\theta_1 \equiv \theta_{\odot}$ and
$\theta_2 \equiv \theta_{{\rm atm}}$, we use the best
fit values found in the analyzes
\cite{fogli,SKatm00} of 
the solar and atmospheric neutrino
data, respectively: 
$\tan^2\theta_{\odot} = 0.46$
and $\sin^22\theta_{{\rm atm}} = 1$.
The angle $\theta_3$ is varied within its allowed
$3\sigma$ range, $\sin^2\theta_3 < 0.05$, while the 
three $CP$ violating phases $\delta$, $\alpha$ and 
$\beta$ are treated as free parameters.
For the normal hierarchical 
neutrino mass spectrum one has: 
$m_1 \ll m_2 \simeq \sqrt{\Delta m^2_{\odot}}
\ll m_3 \simeq \sqrt{\Delta m^2_{{\rm atm}}}$,
where $\Delta m^2_{\odot} > 0$ and 
$\Delta m^2_{{\rm atm}} > 0$ drive 
the solar and atmospheric neutrino oscillations.
In the numerical estimates which follow we use 
$\Delta m^2_{\odot} = 7.3\times 10^{-5}~{\rm eV^2}$
\cite{fogli} and 
$\Delta m^2_{{\rm atm}} = 
2.7\times 10^{-3}~{\rm eV^2}$ \cite{SKatm00}.\\ 
 
   The expression for $m_\nu^\mathrm{th}$ 
and the analysis of interest simplify considerably
if the hierarchy between the masses 
$M_1$ and $M_2$ is relatively mild:
\begin{equation} 
M_2 \simeq 10 \, M_1~.
\label{M12hier}
\end{equation}
%
It is not difficult to convince oneself 
that if Eq.\ (\ref{M12hier}) holds and if in addition --- in agreement 
with the hierarchy requirement in $m_D$ --- the relation  
\begin{equation} \label{eq:mdhie}
m_{D3} \, s_{2R(L)} \simeq 10^{-2} \, m_{D3} \simeq m_{D2} 
\end{equation}
%
holds, the form of $m_\nu^\mathrm{th}$ 
leads to $\sin^2 2 \theta_{2} \equiv
\sin^22\theta_{{\rm atm}} \simeq 1$
without any further fine--tuning as long as the terms
$\propto 1/M_3$ in Eq.\ (\ref{mnuthee}) 
give a sub--leading contribution to $m_\nu^\mathrm{th}$. 
This is ensured provided the inequality
\begin{equation}
M_3 \gg s_{2R}^{-2} \, M_2 \sim 10^{4} \, M_2~,
\label{M23hier}
\end{equation}
%
\noindent implying a strong hierarchy between 
$M_2$ and $M_3$, holds. The existence of a mild
hierarchy between $M_1$ and $M_2$, Eq. (\ref{M12hier}),
and of a strong hierarchy between
$M_2$ and $M_3$, Eq.\ (\ref{M23hier}) is,
as we will see, also compatible with
the requirement of effective leptogenesis.\\ 

  Under the conditions (\ref{M12hier}) --- (\ref{M23hier}) the matrix 
$m_\nu^\mathrm{th}$ takes a relatively simple form:
\begin{equation}
\label{mnuthee}
\ba  \D 
(m_\nu^\mathrm{th})_{ee} \simeq \frac{(m_{D1}- s_{1L} s_{1R} m_{D2} 
e^{ i \alpha_{W}})^{2}}{M_{1}} + \frac{s_{1L}^{2} m_{D2}^{2} 
e^{2 i (\alpha_{W}+\alpha_{R})}}{M_{2}}\\[0.3cm] \D
(m_\nu^\mathrm{th})_{e \mu} \simeq
-\frac{(m_{D1}- s_{1L} s_{1R} m_{D2} e^{i \alpha_{W}}) 
s_{1R}m_{D2} e^{i \alpha_{W}}}{M_{1}}+ 
\frac{m_{D2}^{2} s_{1L}e^{2 i(\alpha_{W}+\alpha_{R})}}{M_{2}}\\[0.3cm] \D
(m_\nu^\mathrm{th})_{e \tau} \simeq
\frac{(m_{D1}- s_{1L} s_{1R} m_{D2} e^{i \alpha_{W}}) 
(s_{1L} s_{2R} - s_{3R} e^{i  \delta_{R}}) m_{D3}^{2} 
e^{2i (\beta_{W}+\delta_{L})}}{M_{1}}- \\[0.3cm] 
\D \frac{m_{D2} m_{D3} s_{1L} s_{2R}e^{ i (\alpha_{W} +\beta_{W}+\delta_{L} 
+ 2 \alpha_{R})}}{M_{2}} \\[0.3cm] \D
%
(m_\nu^\mathrm{th})_{\mu \mu}  
\simeq - \frac{m_{D2}^2}{M_2} \, e^{2 i (\alpha_R + \alpha_W)}~,\\[0.3cm] \D
(m_\nu^\mathrm{th})_{\mu \tau} \simeq 
\frac{m_{D2} \, m_{D3}}{M_2} \, s_{2R} \,
e^{i (2\alpha_R + \alpha_W + \beta_W + \delta_L)}~,\\[0.3cm]\D
 (m_\nu^\mathrm{th})_{\tau \tau} \simeq - \frac{m_{D3}^2}{M_2} \, 
s_{2R}^2 \, e^{2i(\alpha_R + \beta_W + \delta_L)}~,
\ea 
\end{equation}
%
\noindent where we have not written explicitly the sub--leading terms 
$\propto 1/M_1$ and 
$\propto  1/M_3$. 
Further analysis shows that 
for the generic values of the parameters
$m_{Di}$ and $s_{jL(R)}$ 
we use, the matrix $m_\nu^\mathrm{th}$
thus obtained leads to a value of
$\theta_{\odot} \equiv \theta_1$
compatible with the observations
for, e.g., $M_2 \sim 10^{10}$ GeV.
The value of $\theta_{\odot}$ is 
particularly sensitive to the specific
value of $m_{D1}$ chosen: a very 
good agreement with the best fit value
$\tan^2\theta_{\odot} \cong (0.42 - 0.46)$
determined in the
analyses of the solar neutrino data
is obtained for $m_{D1} \cong (70 - 90)$ MeV\@. 

  In what regards the angle $\theta_3$,
for $m_{D1} \sim 100$ MeV and
$s_{1L} \simeq (0.10 - 0.25)$ and 
the standard values of the
other parameters we use, 
one generically has:
$\sin\theta_3 \sim s_{1L}/\sqrt{2}$.  Consequently, 
$\sin^2\theta_3$ is relatively large:
typically one has $\sin^2\theta_3 \gs 0.005$. 
However, it is also possible to fine--tune the values of the
parameters involved in the analysis to get  $\sin^2\theta_{3} < 0.005$ 
and this latter possibility cannot be ruled out.\\

   The above analysis shows that 
the assumptions 
we made about the magnitude of the angles in 
$U_L$ and $U_R$ and of the values of $m_{Di}$
are very well in agreement with the 
existing data. 
We find, in particular, that the angle 
$\theta_3$ can be relatively large 
($\sin^2\theta_{3} \gtap 0.005 - 0.010$),
which has important implications
for the searches for $CP$ violation in 
neutrino oscillations. In fact, we shall see 
that under the assumptions made, 
the $CP$ violating 
observables in neutrino 
oscillations can be sizable.


\subsection{\label{sec:leptobiun}
Leptogenesis in the bi--Unitary Parametrization}


 In the context of leptogenesis
the lepton asymmetry depends on the parameter $m_D^\dagger m_D$, as 
discussed in Section \ref{sec:frame}. 
Using Eq.\ (\ref{eq:mddmd}) we 
find for $(m_D^\dagger m_D)_{11}$ 
to leading order:  
\begin{equation} \label{eq:mdmd11}
\begin{array}{c} \D
(m_D^\dagger m_D)_{11} \simeq 
m_{D2}^2 \, s_{1R}^2 + m_{D3}^2 \left( s_{1R}^2 \, s_{2R}^2 - 
2\, \cos \delta_R \, s_{1R} \, s_{2R} \, s_{3R} + s_{3R}^2 
\right) 
\\[0.3cm]
\D \simeq  \left(m_{D2}^2  + m_{D3}^2 \, s_{2R}^2 \right) \, s_{1R}^2  ~. 
\end{array}
\end{equation}
%
The decay asymmetry $\varepsilon_1$ 
receives contribution from 
\begin{equation} \label{eq:im12}
\begin{array}{c} \D 
{\rm Im}\left\{(m_D^\dagger m_D)_{12}^2 \right\} \simeq  \\[0.3cm]
 2 \left( \cos \alpha_R \,  s_{1R}  \left(m_{D2}^2  + m_{D3}^2 \, 
s_{2R}^2 \right)  -  
m_{D3}^2  \cos(\alpha_R - \delta_R ) \,  s_{2R} \, s_{3R}
\right)  \\[0.3cm]
\D \left( \sin \alpha_R \,  s_{1R}  \left(m_{D2}^2  + m_{D3}^2 \, 
s_{2R}^2 \right)  -  
m_{D3}^2  \sin(\alpha_R - \delta_R ) \,  s_{2R} \, s_{3R}
\right)  \\[0.3cm]
\D \simeq 
\left(m_{D2}^2  + m_{D3}^2 \, s_{2R}^2 \right)^2 
\, s_{1R}^2 \, \sin 2 \alpha_R~, 
\end{array}
\end{equation}
as well as from 
\begin{equation} \label{eq:im13}
\begin{array}{c}
{\rm Im}\left\{(m_D^\dagger m_D)_{13}^2 \right\} \simeq  \\[0.3cm]
2 \, m_{D3}^4 
\left( 
\cos(\beta_R + \delta_R) \, s_{1R} \, s_{2R} - \cos \beta_R \, s_{3R}
\right) \left(
\sin(\beta_R + \delta_R) \, s_{1R} \, s_{2R} - \sin \beta_R \, s_{3R}
\right)\\[0.3cm]
\simeq m_{D3}^4 \,  s_{1R}^2 \, s_{2R}^2 \, \sin 2(\beta_R + \delta_R)~, 
\end{array}
\end{equation}
where $s_{3R}$ was neglected in the last 
expression. 
Assuming a hierarchical mass spectrum of the heavy 
Majorana neutrinos one finds for the decay asymmetry  
\begin{equation} \label{eq:epsapp}
\varepsilon_1 \simeq -\frac{3}{2} \, \frac{1}{8 \, \pi \, v^2} \, 
\left( ( m_{D2}^2  + m_{D3}^2 \, s_{2R}^2) \, \sin 2 \alpha_R \,  
\frac{M_{1}}{M_{2}} + 
\frac{m_{D3}^4 \, s_{2R}^2}{m_{D2}^2  + m_{D3}^2 \, s_{2R}^2} \,  
\sin 2( \beta_R + \delta_R) \, \frac{M_{1}}{M_{3}} 
\right)~.
\end{equation}
%
Therefore, we can identify $\alpha_R$ 
and $(\beta_R + \delta_R)$ 
as the leptogenesis phases. Note that, 
as it should, $\varepsilon_1 = 0$ 
for $\alpha_R, \beta_R, \delta_R = 0, \pi/2, \pi$.  
Taking $m_{D2} \simeq 10^{-2} \, m_{D3}$
and $s_{2R} \simeq 10^{-2}$,
we can estimate the decay asymmetry as 
\footnote{Note that the form of this 
equation resembles Eq. (17) 
in the second paper in Ref.\ \cite{ich}. 
A type II see--saw mechanism was used in that work.}   
\begin{equation} \label{eq:epsapp1}
\varepsilon_1 \sim - 10^{-9} \left(\frac{m_{D3}}{\rm GeV} \right)^2 
\, \left( \sin 2 \alpha_R \, \frac{M_{1}}{M_{2}} 
+ 10^{3} \, \sin 2( \beta_R + \delta_R) \, \frac{M_{1}}{M_{3}} \right)~. 
\end{equation}
%
As discussed earlier, 
the heavy Majorana neutrino 
masses $M_{i}$ are also expected to posses a 
hierarchy. Under the condition (\ref{M23hier})
of a strong hierarchy between $M_2$ and $M_3$,
the second term in the last equation is negligible, 
and only the phase $\alpha_R$ is relevant for leptogenesis.
If further one has $m_{D3} \sim 10^2$ GeV,
as is suggested by up--quark mass values,
it is possible to have
$\varepsilon_1 \sim 10^{-6} - 10^{-7}$ 
if, e.g., $M_{1}/M_{2} \sim 0.1$,  
which is compatible with the requirement
of a mild hierarchy between $M_1$ and $M_2$,
Eq.\ (\ref{M12hier})  which in the previous Section 
was shown to be also compatible with the low energy neutrino 
phenomenology.\\

Identifying $m_D$ with the 
down--quark or charged lepton mass matrix, 
i.e., $m_{D3} \sim 1 $ GeV, 
leads to a rather small value of 
$\varepsilon_1$ for 
$M_1 \ll M_2 \ll M_3$. In this case 
the lepton asymmetry 
can only be amplified 
to the requisite value by the 
resonance mechanism \cite{self}
which is operative for quasi--degenerate 
heavy Majorana neutrinos. Thus, 
leptogenesis with hierarchical neutrino Majorana and 
Dirac masses prefers up--quark type masses for the latter, 
a fact first noticed in \cite{first}.\\

  The ``effective mass'' $\tilde{m}_1$, 
that represents a sensitive parameter 
for the Boltzmann equations, is given by 
\begin{equation}
\tilde{m}_1 = \frac{(m_D^\dagger m_D)_{11}}{M_{1}} 
\sim \left(\frac{m_{D3}}{\rm GeV} \right)^2 \, 
\left(\frac{10^{9} \rm GeV}{M_{1}} \right) 10^{-5} \, \rm eV ~. 
\end{equation}
%
For $m_{D3} \sim 10^2$ GeV and 
$M_{1}$ of $10^9$ to $10^{10}$ GeV, acceptable values of 
$10^{-1}$ to $10^{-3}$ eV for $\tilde{m}_1$ are obtained.
This leads to values of $\kappa \sim 10^{-1} - 10^{-3}$. 
Assuming that the sine of the 
phases in Eq.\ (\ref{eq:epsapp}) is not small, 
a value of $Y_B \sim 10^{-10}$ 
is therefore ``naturally'' obtained 
for, e.g., $m_{D3} \sim 10^2$ GeV, $M_{1} \sim 10^9$ GeV, and 
$M_{2} \sim 10^{10} \, {\rm GeV} \ll M_{3} \sim 10^{15}$ GeV. 
The indicated values of $M_1$, $M_2$ and $M_3$ 
are also in very good agreement with the constraints
Eqs.\ (\ref{M12hier}) and (\ref{M23hier}).  


\subsection{\label{sec:muegbiun}LFV Charged Lepton Decays} 

 
According to Eq.\ (\ref{eq:muegBR}), the 
branching ratios of the LFV charged 
lepton decays, $BR(\ell_i \rightarrow \ell_j + \gamma)$,
depend on  $(m_D L m_D^\dagger)$. 
We have calculated this matrix 
with the usual approximations for the masses and 
angles and found the leading terms to be 
\begin{equation}
\begin{array}{cc}
(m_D L m_D^\dagger)_{21} \simeq & m_{D2}^2 \, s_{1L} \, L_2, \\[0.4cm] 
(m_D L m_D^\dagger)_{31} \simeq & m_{D3}^2 \, (s_{1L}s_{2L} - s_{3L} \, 
e^{i \delta_L}) \, L_3 + m_{D3}m_{D2}s_{1L}s_{2R} (L_2 - L_3). \\[0.4cm] 
(m_D L m_D^\dagger)_{32} \simeq & m_{D3}^2 \, s_{2L} \, L_3. 
\end{array}
\end{equation}
%
Barring accidental cancellations in 
the expression for $(m_D L m_D^\dagger)_{31}$,
we obtain the following general relation
between the branching ratios
of interest
\footnote{Obviously, the relation we obtain 
does not hold if, e.g., the accidental cancellation 
$(s_{1L}s_{2L} - s_{3L} e^{i \delta_L}) = 0$
takes place. The latter requires, 
however, a fine--tuning of the values of
4 parameters.}:  
\begin{equation}
BR(\tau \rightarrow \mu + \gamma) \gg  
BR(\tau \rightarrow e + \gamma) \gg 
BR(\mu \rightarrow e + \gamma) ~.   
\end{equation}
%
\noindent If we use the generic
values of the relevant parameters we
work with, we find
\begin{equation}
BR(\tau \rightarrow \mu + \gamma) \sim 
10^2 \, BR(\tau \rightarrow e + \gamma) \sim 
 10^5 \, BR(\mu \rightarrow e + \gamma) ~.   
\label{LVFBR}
\end{equation}
%
  
   The results do not differ 
significantly from the results obtained 
by using a common mass scale $M$ and taking the log--terms out of the 
matrix, i.e., for $ (m_D m_D^\dagger) \, \log M_X/M$ 
instead of $m_D L m_D^\dagger$. 
Note that though $m_D$ contains 
$U_L$ and $U_R$, only the angles and phases 
of $U_L$ appear in $BR(\ell_i \rightarrow \ell_j + \gamma)$,
which confirms that Eq.\ (\ref{eq:mdmdd}) 
represents a rather good 
approximation for estimating  
$BR(\ell_i \rightarrow \ell_j + \gamma)$.

\section{\label{sec:connection?}Leptogenesis, \betabeta--decay and 
Low Energy Leptonic $CP$ Violation} 

We shall investigate next 
whether there exists a relation between
the high energy $CP$ violation leading to 
the decay asymmetry Eq.\ (\ref{eq:epsapp})  
and low energy observables. We will consider the effective 
Majorana mass in \betabeta--decay and the $CP$ violating asymmetry in 
neutrino oscillations. 


\subsection{\label{sec:meff}  The Effective 
Majorana Mass in \betabeta--Decay}  


The observation of \betabeta--decay  would provide direct evidence 
for the non--conservation of the total lepton charge, which is 
a necessary ingredient for the leptogenesis mechanism.  
In \betabeta--decay one measures
the absolute value of the 
$ee$ element of $m_\nu$. Given the 
form of $m_D$ in Eq.\ (\ref{eq:mdapp}), 
it is not difficult to find this element:  
\begin{equation} \label{eq:meffapp}
\meff \simeq  \frac{(m_{D1} - s_{1L} s_{1R} m_{D2}
e^{i \alpha_W})^{2}}{M_{1}}
+
\frac{m_{D2}^{2} e^{2i (\alpha_W + \alpha_{R})}}{M_{2}} 
+\frac{m_{D3}^2 \, s_{3L}^2 \, 
e^{2i (\beta_R + \beta_W + \delta_R)}}{M_3}~. 
\end{equation}
%
The term of order $m_{D3}^2$ contains, in particular, the parameters 
$\alpha_W$ and $\beta_{W}$, which do not influence 
the baryon asymmetry.  
The condition that this term does 
not contribute significantly to $|\meff|$ is 
\begin{equation} \label{eq:meffdec}
\baz 
\D \frac{m_{D3}^2}{M_{3}} \, s_{3L}^2 \ll \left\{
\ba 
\D \frac{m_{D2}^2}{M_{1}} \, s_{1L}^2 \, s_{1R}^2 \\[0.4cm] 
\D \frac{m_{D2}^2}{M_{2}} \, s_{1L}^2 
\ea \right. 
& 
\D \Rightarrow M_{3} \gg \left\{
\ba 
\D M_{1} \, \left( \frac{m_{D3}}{m_{D2}} \right)^2 \, 
\frac{s_{3L}^2}{s_{1L}^2 \, s_{1R}^2} \sim 10^2 \, M_{1} \\[0.4cm]
\D M_{2} \, \left( \frac{m_{D3}}{m_{D2}} \right)^2 \,
\frac{s_{3L}^2}{s_{1L}^2 } \sim M_{2} 
\ea \right.
\ea , 
\end{equation}
%
which is compatible with Eqs. (24) and (26) and
the hierarchy between $M_3$ and $M_{1,2}$,  
required for generating a sufficient baryon asymmetry (see the 
discussion after Eq.\ (\ref{eq:epsapp1})). 
The first two terms in (\ref{eq:meffapp}) can 
contribute comparably to $|\meff|$ unless 
$M_{2} \, s_{1R}^2 \gg M_{1}$ or $M_{2} \, s_{1R}^2 \ll M_{1}$. 
From the two main possibilities for $m_D$, 
the choice of the  up--quark type matrix,
$m_D \sim m_{\rm up}$, favored by leptogenesis
in the case of hierarchical
$M_i$ and $m_D$,
leads to $|\meff{}|$   
which is larger by a 
factor of $(m_c/m_{s, \mu})^2 \sim 10^2$
then in the case of $m_D \sim m_{\rm down}$,
$m_c~(m_s)$ being the charm (strange) quark mass.
If $m_{D2} \sim 1$ GeV and 
$s_{1L(R)}^2 \sim (10^{-2} - 10^{-1})$ 
(see Eq.\ (25)), one finds for  
$M_{1} \sim 10^{9}$ GeV and 
$M_{2} \sim 10^{10}$ GeV 
that $|\meff| \sim (0.001 - 0.01)$ eV
 \footnote{We would like to emphasize that the 
 numerical values for the observables 
we give should be taken 
 not too literally: a spread within 
 one order of magnitude, as included in the prediction for 
 $|\meff|$ we give, should be 
 allowed for.},  
which may be within the reach 
of the 10t version of the
GENIUS experiment \cite{0vbbrev}.\\
 
If condition (\ref{eq:meffdec}) holds, 
only the leptogenesis parameter 
$\alpha_R$ appears in the effective Majorana mass
$|\meff|$. This condition corresponds to 
a rather strong hierarchy between 
the masses of the $N_i$ and thus to 
a decoupling of the heaviest Majorana neutrino $N_3$.   
In this case, 
one sees from Eq.\ (\ref{eq:epsapp}) that only the 
term proportional to $\sin 2 \alpha_R$ contributes 
to $Y_B$ and there is 
a direct correlation 
between the rate of \betabeta--decay and the 
baryon asymmetry of the Universe. 
For a mild hierarchy between the masses 
$M_i$, the unknown phases $\alpha_W$ and $\beta_W$ spoil any 
simple connection between $|\meff|$ and $Y_B$.\\


\subsection{\label{JCP} 
$CP$ Violation in Leptogenesis and
in Neutrino Oscillations} 


  Manifest $CP$ violation can be probed in 
neutrino oscillation experiments (see, e.g., \cite{CPosc}).
The corresponding $CP$ violating observables
depend on the rephasing invariant quantity \cite{PKSP88}
$J_{CP}$ --- the leptonic analog of the Jarlskog invariant.
The following form \cite{lissac} of $J_{CP}$ is particularly suited for 
our analysis: 
\begin{equation}
J_{CP} = -\frac{{\rm Im} (h_{12} \, h_{23} \, h_{31} )}
{\Delta m^2_{21} \, \Delta m^2_{31} \, \Delta m^2_{32}}
~,  \mbox{ where } 
h = m_\nu m_\nu^\dagger ~. 
\end{equation}
%
The invariant $J_{CP}$ can be calculated using
Eqs.\ (\ref{eq:seesaw}) and (\ref{eq:mdapp}). 
The resulting expression 
is rather long and which is the 
leading term depends on the degree of 
hierarchy of the heavy Majorana neutrino masses. 
In general, however, we find that $J_{CP}$ vanishes for 
$m_{D2} = 0$ and/or $1/M_{3} = 0$. These approximations  
correspond effectively to a 2 flavour neutrino mixing case,
and consequently to an absence of Dirac--like $CP$ 
violation. In general, all six phases contribute 
to $J_{CP}$, as is suggested by Eq.\ (\ref{eq:mnupara}).\\ 

   As an example, let us consider the hierarchy
$M_{1} \simeq 10^{-1} \, M_{2} \simeq 10^{-4} M_{3}$. Then 
there are four leading terms in $J_{CP}$, which are proportional to 
$(M_{1} \, M_{2}^4 \, M_{3})^{-1 }$, $(M_1 \, M_3)^{-3}$, 
$(M_{1} \, M_{2} \, M_{3}^4)^{-1 }$ and 
$(M_{1} \, M_{2} \, M_{3})^{-2}$, respectively. They read: 
\begin{equation} \label{eq:Japp}
\D \ba
{\rm Im} (h_{12} \, h_{23} \, h_{31} ) \simeq 
\frac{\D -2 \, m_{D2}^5 \, 
m_{D3}^7 \, s_{1L}^2 \, s_{1R}^2 \, s_{2L} \, s_{2R}^5}
{\D M_{1} \, M_{2}^4 \, M_{3}} \, 
\cos 2\alpha_R 
\sin \left(\alpha_W -( \beta_W + \delta_L ) + 2 \alpha_R - 
2 (\beta_R + \delta_R)   \right)\\[0.3cm]
- \frac{\D  \, m_{D2}^5 \, 
m_{D3}^7 \, s_{1L}^2 \, s_{1R}^5 \, s_{2L} \, s_{3R}}
{\D M_{1}^3 \, M_{3}^3} \, 
\sin \left(\alpha_W - 2 \, \beta_R - \beta_W - \delta_L  - \delta_R 
\right)\\[0.3cm] 
+ \frac{\D  \, 2 \, m_{D2}^4 \, 
m_{D3}^8 \, s_{1L}^2 \, s_{1R}^2 \, s_{2L} \, s_{3L}} 
{\D M_{1} \, M_2 \, M_{3}^4} \, \cos 2 \alpha_R \sin \delta_L \\[0.3cm]
+ \frac{\D  \, m_{D2}^7 \, 
m_{D3}^5 \, s_{1L}^2 \, s_{1R}^4 \, s_{2L} \, s_{2R}} 
{\D M_{1}^2 \, M_2^2 \, M_{3}^2} \, \left( 
2\, \sin \left(\alpha_W - \beta_W - \delta_L \right) + 
\sin \left( \alpha_W - 4\, \alpha_R - \beta_W - \delta_L \right) 
\right)\\[0.3cm] 
+ \frac{\D  \, m_{D2}^5 \, 
m_{D3}^5 \, s_{1L}^2 \, s_{1R}^4 \, s_{2L} \, s_{2R} }  
{\D M_{1}^2 \, M_2^2 \, M_{3}^2} \, \left( m_{D2}^2 \, \left( 
2\, \sin \left(\alpha_W - \beta_W - \delta_L \right) + 
\sin \left( \alpha_W - 4\, \alpha_R - \beta_W - \delta_L \right) 
\right) \right. \\[0.3cm]
\left. + m_{D3}^2 \, s_{2R}^2 \, \sin \left(\alpha_W - \beta_W - \delta_L 
\right) \right)~,  
\ea 
\end{equation}
%
which illustrates our general conclusions mentioned above. 
Choosing a different hierarchy of the heavy Majorana neutrino masses 
will lead to the presence of different terms with different combinations 
of the parameters, especially of the phases.\\   

 Of the 6 independent physical $CP$ violating phases
of the see--saw model, 5 phases are present in 
Eq.\ (\ref{eq:Japp}), namely 
$\alpha_R$, $\alpha_W - \beta_W - \delta_L$, $\beta_R$, $\delta_R$ 
and $\delta_L$. This can be understood 
as being due to the fact that in the approximations we use 
$m_{D1}$ is neglected. 
Setting in addition $s_{3L(R)}$ to zero 
should remove two more phases from the parameter space. 
Indeed, one finds from Eq.\ (\ref{eq:Japp}) 
that only three independent 
phases ($\alpha_R$, $\alpha_W - \beta_W - \delta_L$ and $\beta_R + \delta_R$) 
enter into the expression for $J_{CP}$ in this case. 
Order--of--magnitude--wise we can predict the magnitude of the 
$CP$ violation to be 
\begin{equation}
J_{CP} \sim 10^{-2} \, 
\left(\frac{10^{9} \rm \, GeV}{M_{1}} \right) \, 
\left(\frac{10^{10} \rm \, GeV}{M_{2}} \right)^4 \, 
\left(\frac{10^{14} \rm \, GeV}{M_{3}} \right) \, 
\sin \big(\alpha_W -( \beta_W + \delta_L ) + 
2 \alpha_R - 2 (\beta_R + \delta_R)   \big)~.  
\end{equation}
%
 One sees that within the approximation discussed in this paper 
there is no connection, in general, between 
leptogenesis in Eq.\ (\ref{eq:epsapp1}) and the low energy $CP$ 
violation in neutrino oscillations. 
Successful leptogenesis without low energy $CP$ 
violation in neutrino oscillations is, in principle, possible. 
This interesting case has recently been 
discussed in \cite{nolowCP}. However, given the results of 
the present Subsection, fine--tuning between 
the values of the different 
phases has to take place in order to have 
leptogenesis and $J_{CP} \approx 0$. 
Even if there is no direct connection between the leptogenesis and 
the value of $J_{CP}$, a measurement of $J_{CP}$ can shed light
on the ``non--leptogenesis'' parameters.  
We note finally that the decoupling of leptogenesis 
from the low energy  
Dirac--phase has been noticed to take place 
in a number of models
as well, e.g.\ in \cite{others2}.\\ 

    We comment finally on an interesting 
special case of Eq.\ (\ref{eq:Japp}). 
Suppose that the phases ``conspire'' 
to fulfill the relations 
$ \alpha_W - \beta_W - \delta_L = \beta_R = \delta_R = \delta_L = 0$ or, 
for $s_{3L(R)} = 0$, 
$\alpha_W - \beta_W - \delta_L = \beta_R + \delta_R = 0$. 
Then, only one phase contributes to the $J_{CP}$ asymmetry and 
one finds from Eq.\ (\ref{eq:Japp}) that 
$J_{CP} \propto \sin 4 \alpha_R$. 
The baryon asymmetry, on the other hand, 
is proportional to $\sin 2 \alpha_R$ as can be seen from Eqs.\ 
(\ref{eq:YB}) and (\ref{eq:epsapp}). This situation allows for a 
correlation between the relative sign of the 
baryon asymmetry of the Universe and 
the $CP$ asymmetry in neutrino oscillations. 
The two quantities posses opposite signs 
if $2\alpha_R$ lies between $\pi/2$ and $\pi$ 
and the same signs for values 
of  $2\alpha_R$ outside this range. 
The parameter space allowing for 
this correlation
corresponds to the 
$3 \times 2$ see--saw model 
with 2 texture zeros in 
the Dirac mass matrix discussed in Ref.\ \cite{23}. 


\section{\label{sec:concl}Conclusions}


  Assuming only a hierarchical structure of the heavy 
Majorana neutrino masses $M_{1,2,3}$,
$M_1 \ll M_2 \ll M_3$,
and of the Dirac mass matrix 
$m_D$ of the see--saw mechanism (see Eq.\ (\ref{eq:mdass})),
and working in the bi--unitary 
parametrization of $m_D$, we find that 
in order to produce a sufficient amount of 
baryon asymmetry via the leptogenesis  
mechanism, the scale of $m_D$ should be given by the  
up--quark masses. In this class of 
``hierarchical'' see--saw
models, the branching ratios 
of LFV charged lepton decays, 
whose dependence on $m_D$ is 
introduced by RGE effects 
within the SUSY GUT version of the model, 
are predicted to fulfill the relations  
$BR(\tau \rightarrow \mu + \gamma) \gg 
BR(\tau \rightarrow e + \gamma) 
\gg BR(\mu \rightarrow e + \gamma)$: typically one has
$BR(\tau \rightarrow \mu + \gamma) \sim 10^2 \, 
BR(\tau \rightarrow e + \gamma) 
\sim 10^5 \, BR(\mu \rightarrow e + \gamma)$. 
We find that the effective Majorana mass
in $\betabeta$--decay 
depends on the $CP$ violating phase 
controlling the leptogenesis 
if one of the heavy Majorana neutrinos 
is much heavier than the other two,
$M_3 \gg 10^4 \, M_2$. 
A rather mild hierarchy between 
the masses of the lighter two heavy 
Majorana neutrinos, $M_2 \sim 10 \, M_1$,
is required for successful leptogenesis. 
The hierarchical relations 
$M_3 \gg 10^4 \, M_2$ and
$M_2 \sim 10 \, M_1$ with, e.g.,
$M_1 \sim 10^{9}$ GeV, are also compatible
with the low--energy neutrino mixing phenomenology.
The $CP$ violation effects in neutrino 
oscillations can be observable.
In general, there is no direct connection 
between the latter and the $CP$ violation
in leptogenesis.
We find, however, that if 
the $CP$ violating phases
of the see--saw model ``conspire'' to satisfy 
certain relations, the baryon asymmetry 
of the Universe and the leptonic 
$CP$ violation rephasing invariant
$J_{CP}$, which determines 
the magnitude of the $CP$ violation
effects in neutrino oscillations,
depend on the same $CP$ violating phase and
their signs are correlated. 

\vspace{0.5cm}
\leftline{{\bf Acknowledgments}}
This work was supported in part by 
the EC network HPRN-CT-2000-00152
(S.T.P.\ and W.R.), by the Italian 
MIUR under the program ``Fenomenologia delle 
Interazioni Fondamentali'' (S.T.P.),
and by the US Department of Energy Grant DE-FG03-91ER40662
(S.P.). S.P.\ would like to thank the Elementary Particle 
Physics Sector
of SISSA (Trieste, Italy) and the Inst.\ de Fisica Teorica
of the Universidad Autonoma de Madrid (Spain)
for kind hospitality 
during part of this study.

\newpage

\end{document}